\documentclass[preprintnumbers,amsmath,amssymb,pre,floatfix,showkeys,superscriptaddress,nofootinbib,secnumarabic]{revtex4}

\usepackage[english]{babel}
\usepackage[latin2]{inputenc}
\usepackage[T1]{fontenc}
\usepackage{color}
\usepackage{amsmath}
\usepackage{amssymb}
\usepackage{ae,aecompl}
\usepackage{enumerate}
\usepackage{epsfig}
\usepackage{graphics}
\usepackage{graphicx}
\usepackage{pifont}
\usepackage[sort&compress]{natbib}
\usepackage{wasysym}
\usepackage{rotating}
\usepackage{multirow}

\newcommand{\addsmallfig}[4]
{
\begin{figure}[!ht]
\centerline{\includegraphics[width=155pt,angle=-90]{#2#3}}
\caption{#4}
\label{fig:#1}
\end{figure}
}

\newcommand{\addwidefig}[4]
{
\begin{figure*}[!tp]
\centerline{\includegraphics[width=170pt,angle=-90,trim=180 0 50 30]{#2#3}}
\caption{#4}
\label{fig:#1}
\end{figure*}
}

\newcommand{\addtwofigs}[5]
{
\begin{figure*}[!t]
\centerline{\includegraphics[height=220pt,angle=-90]{#2#3}\includegraphics[height=220pt,angle=-90]{#2#4}}
\caption{#5}
\label{fig:#1}
\end{figure*}
}

\catcode`\_=12\catcode`\_=8

\def\ev #1{\left\langle #1 \right\rangle}

\def\MO{\mathrm{MO}}
\def\CA{\mathrm{CA}}
\def\LO{\mathrm{LO}}
\def\MOO{\mathrm{MO}^0}
\def\CAO{\mathrm{CA}^0}
\def\LOO{\mathrm{LO}^0}
\def\R{{\cal R}}
\def\DeltaR{\Delta^\mathrm{R}}

\def\ccc#1;#2{\left\langle #1 \left\vert #2 \right.\right\rangle}
\def\ev #1{\left\langle #1 \right\rangle}

\begin{document}

\preprint{}
\title{Models for the impact of all order book events}
\author{Zolt\'an Eisler}
\affiliation{Capital Fund Management, Paris, France}
\author{Jean-Philippe Bouchaud}
\affiliation{Capital Fund Management, Paris, France}
\author{Julien Kockelkoren}
\affiliation{Capital Fund Management, Paris, France}

\date{\today}

\begin{abstract}
We propose a general framework to describe the impact of different events in the order book, that generalizes previous work on the impact of market orders. 
Two different modeling routes can be considered, which are equivalent when only market orders are taken into account. One model posits that each event type
has a temporary impact (TIM). The ``history dependent impact'' model (HDIM), on the other hand, assumes that only price-changing events have a direct impact, 
itself modulated by the past history of all events through an ``influence matrix'' that measures how much, on average, an event of a given type affects 
the immediate impact of a price-changing event of the same sign in the future. We find in particular that aggressive market orders tend to reduce the impact of 
further aggressive market orders of the same sign (and increase the impact of aggressive market orders of opposite sign). We discuss the relative merits of TIM and 
HDIM, in particular concerning their ability to reproduce accurately the price diffusion pattern. We find that in spite of theoretical inconsistencies, TIM appears to 
fare better than HDIM when compared to empirical data. We ascribe this paradox to an uncontrolled approximation used to calibrate HDIMs, calling for further work 
on this issue.
\end{abstract}


\maketitle

\tableofcontents


\section{Introduction}

The relation between order flow and price changes has attracted considerable attention in the recent years \cite{hasbrouck.book, mike.empirical, bouchaud.subtle, bouchaud.molasses, lyons.book, bouchaud.review}. Most empirical studies to date have focused on the impact of (buy/sell) market orders. Many interesting results have been obtained, such as the very weak dependence of impact on the volume of the market order, the long-range nature of the {\it sign} of the trades, and the resulting non-permanent, power-law decay of market order impact with time (see Ref. \cite{bouchaud.review}). However, this representation of impact is incomplete in at least two ways. First, the impact of all market orders is usually treated on the same footing, or with a weak dependence on volume, whereas some market orders are ``aggressive'' and immediately change the price, while other are more passive and only have a delayed impact on the price. Second, other types of order book events (limit orders, cancellations) must also directly impact prices: adding a buy limit order induces extra upwards pressure, and cancelling a buy limit order decreases this pressure. Within a description based on market orders only, the impact of limit orders and cancellations is included in an indirect way, in fact as an effectively decaying impact of market orders. This decay 
reflects the  ``liquidity refill'' mechanism explained in detail in Refs. \cite{bouchaud.molasses,weber.rosenow,gerig.phd,bouchaud.review,lillo-farmer.qf}, whereby market order trigger a counterbalancing flow of limit orders. 

A framework allowing one to analyze the impact of all order book events, and to understand in detail the statistical properties of price time series, is clearly desirable. 
Surprisingly, however, there are only very few quantitative studies of the impact of limit orders and cancellations \cite{hautsch.limit,eisler.long} -- 
partly due to the fact that more detailed data, beyond trades and quotes, is often needed to conduct such studies. The aim of the present paper is to provide a theoretical framework for the impact of all order book events, which allows one to build statistical models with intuitive and transparent interpretations \cite{eisler.long}.

\section{A short summary of market order impact models}
\label{sec:shortsum}

A simple model relating prices to trades posits that the mid-point price $p_t$ just before trade $t$ can be written as a linear superposition of the 
of the impact of all past trades \cite{bouchaud.subtle, bouchaud.molasses}:
\begin{equation}
    p_t = \sum_{t'<t}\left[{G}(t-t')\xi_{t'} + \eta_{t'} \right] + p_{-\infty}, \qquad \xi_t \equiv \epsilon_{t}v_{t}^\theta
    \label{eq:ptMO}
\end{equation}
where $v_{t'}$ is the volume of the trade at time $t'$, $\epsilon_{t'}$ the sign of that trade ($+$ for a buy, $-$ for a sell), and $\eta_t$ is an
independent noise term that models any price change not induced by trades (e.g. jumps due to news). The exponent $\theta$
is found to be small. The most important object in the above equation is the function ${G}(\ell)$ 
describing the temporal evolution of the impact of a single trade, which can be called a `propagator': how does the impact of the trade at 
time $t'=t-\ell$ propagate, on average, up to time $t$? Because the signs of trades are strongly auto-correlated, ${G}(\ell)$ must decay with time in a very specific way, in order maintain the (statistical) efficiency of prices. Clearly, if ${G}(\ell)$ did not decay at all, the returns\footnote{In the following, we only focus on price changes over small
periods of time, so that an additive model is adequate.} $r_t = p_{t+1} - p_t$ would simply be proportional to the sign of the trades, and therefore would themselves be strongly autocorrelated in time. 
The resulting price dynamics would then be highly predictable, which is not realistic. The result of Ref. \cite{bouchaud.subtle} is that if the correlation of signs $C(\ell) = \langle \epsilon_t \epsilon_{t+\ell} \rangle$ decays at large $\ell$ as $\ell^{-\gamma}$ with $\gamma < 1$ (as found empirically), then ${G}(\ell)$ must decay 
as $\ell^{-\beta}$ with $\beta = (1-\gamma)/2$ for the price to be exactly diffusive at long times. The impact of single trades is therefore predicted to decay as a power-law 
(at least up to a certain time scale). 

The above model can be rewritten in a completely equivalent way in terms of returns, with a slightly different interpretation \cite{gerig.phd,lillo-farmer.qf}:
\begin{equation}
    r_t = G(0)  \xi_t  + \sum_{t'<t} \kappa(t-t')\xi_{t'} + \eta_t; \qquad \kappa(\ell) \equiv {G(\ell+1) - G(\ell)}.
    \label{eq:ptMO2}
\end{equation}
This can be read as saying that the $t^{th}$ trade has a permanent impact on the price, but this impact is history dependent and depends on the sequence of past trades. The fact that  
$G$ decays with $\ell$ implies that the kernel $\kappa$ is negative, and therefore that a past sequence of buy trades ($\xi_{t' < t} > 0$) tends to {\it reduce} the impact of a further 
buy trade, but increase the impact of a sell trade. This is again a consequence of the dynamical nature of liquidity: when trades persist in a given direction, opposing limit orders tend to 
pile up and reduce the average impact of the next trade in the same direction. For the price to be an exact martingale, the quantity $\widehat \xi_t=-\sum_{t'<t} \kappa(t-t')\xi_{t'}$ must be equal to the
conditional expectation of $\xi_t$ at time $t^-$, such that $\xi_t  - \widehat \xi_t$ is the surprise part of $\xi_t$. This condition allows one to recover the above mentioned 
decay of $G(\ell)$ at large $\ell$.

In order to calibrate the model, one can use the empirically observable impact function ${\cal R}(\ell)$, defined as:
\begin{equation}
    {\cal R}(\ell) = \langle (p_{t+\ell}-p_t) \cdot \xi_t \rangle,
\end{equation}
and the time correlation function $C(\ell)$ of the variable $\xi_t=\epsilon_{t}v_{t}^\theta$ to map out, numerically, the complete shape of ${G}(\ell)$. This was done 
in Ref. \cite{bouchaud.molasses}, using the exact relation: 
\begin{equation}
\label{eq:RCG}
    {\cal R}(\ell)  = \sum_{0 < n\leq \ell} {G}(n) C(\ell-n) + \sum_{n>0} [{G}(n+\ell)-G(n)] C(n).
\end{equation}
Alternatively, one can use the `return' version of the model, Eq. \eqref{eq:ptMO2}, which gives:
\begin{equation}
 {\cal S}(\ell) = \langle r_{t+\ell} \cdot \xi_t \rangle,
\end{equation}   
that in turn leads to
\begin{equation}
\label{eq:SCG}
  {\cal S}(\ell)  = G(0) C(\ell) + \sum_{n>-\ell} {\kappa}(\ell+n) C(n).
\end{equation}   
As noted in Ref. \cite{e.serie}, this second implementation is in fact much less sensitive to finite size effects and therefore more adapted to data analysis.\footnote{The key difference is that a numerical solution necessarily truncates $G(\ell)$ in Eq. \eqref{eq:RCG} and $\kappa(\ell)$ in Eq. \eqref{eq:SCG} at some arbitrary $\ell_\mathrm{max}$. This truncation in the former case corresponds to the boundary condition of $\kappa(\ell > \ell_\mathrm{max}) \equiv 0$, hence a fully temporary impact at long times, while in the latter case to $\kappa(\ell > \ell_\mathrm{max}) \equiv G(\ell_\mathrm{max})$, hence a partially permanent impact. The latter solution is more smooth and consequently it is better behaved numerically.} 

The above model, regardless of the type of fitting, is approximate and incomplete in two interrelated ways. First, Eqs. \eqref{eq:ptMO} and \eqref{eq:ptMO2} neglect the fluctuations of the impact: one expects in general that $G$ and $\kappa$ should depend both on $t$ and $t'$ and not only on $\ell=t-t'$. Impact can indeed be quite different depending on the state of the order book and the market conditions at $t'$. As a consequence, if one blindly uses Eqs. \eqref{eq:ptMO} and \eqref{eq:ptMO2} to compute the second moment of the price difference, $D(\ell)=\langle (p_{t+\ell}-p_t)^2 \rangle$, with a non-fluctuating ${G}(\ell)$ calibrated to reproduce the impact function ${\cal R}(\ell)$, the result clearly underestimates the empirical price variance: see Fig. \ref{fig:vol_test_just_MO_smalltick}.

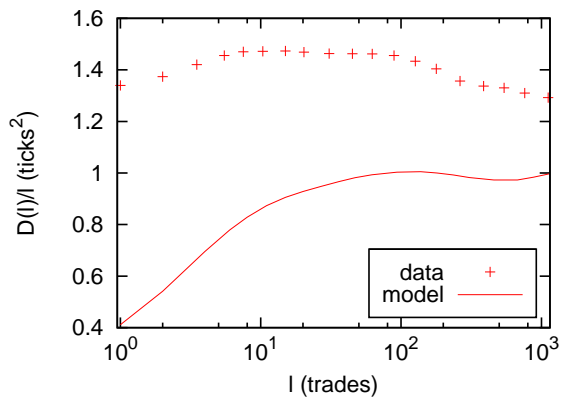
\addsmallfig{vol_test_just_MO_smalltick}{./}{vol_test_just_MO_smalltick}{$D(\ell)/\ell$ and its approximation with the transient impact model 
(TIM) with only trades as events, with $\eta_t=0$ and for small tick stocks. Results are shown when assuming that all trades have the same, non fluctuating impact $G(\ell)$, calibrated to reproduce ${\cal R}(\ell)$. This simple model accounts for $\sim 2/3$ of the long term volatility. Other events and/or the fluctuations of ${G}(\ell)$ must therefore contribute to the market volatility as well.}

Adding a diffusive noise $\eta_t \neq 0$ would only shift $D(\ell)/\ell$ upwards, but this is insufficient to reproduce the empirical data. 
Second, as noted in the introduction, other events of the order book can also change the mid-price, such as limit orders placed inside the bid-ask spread, or cancellations of all the volume at the bid or the ask. These events do indeed contribute to the price volatility and should be explicitly included in the description. A simplified description of price changes in terms of market orders only attempts to describe other events of the order book in an effective way, through the non-trivial time dependence of ${G}(\ell)$. 

In the following, we will generalize the above model to account for the impact of other types of events, beyond market orders. In this case, however, it will become apparent that the two versions 
of the above model {\it are no longer equivalent}, and lead to different quantitative results. Our main objective will be to come up with a simple, intuitive model that (i) can be easily calibrated on
data, (ii) reproduces as closely as possible the second moment of the price difference $D(\ell)$, (iii) can be generalized to richer and richer data sets, where more and more events are observable and 
(iv) can in principle be systematically improved.

\section{Many-event impact models}

\subsection{Notations and definitions}

The dynamics of the order book is rich and complex, and involves the intertwined arrival of many types of events. These events can be categorized in different, more or less natural types. 
In the empirical analysis presented below, we have considered the following six types of events:
\begin{itemize}
\item market orders that do not change the best price (noted $\MOO$) or that do change the price (noted $\MO'$), 
\item limit orders at the current bid or ask ($\LOO$) or inside the bid-ask spread so that they change the price ($\LO'$), 
\item and cancellations at the bid or ask that do not remove all the volume quoted there ($\CAO$) or that do ($\CA'$). 
\end{itemize}
Of course, events deeper in the order book could also be added, as well as any extra division of the above types into subtypes, if more information is available. For example, if the identity code 
of each order is available, one can classify each event according to its origin, as was done in Ref. \cite{toth.brokerage}. The generic notation for an event type occurring at event time $t$ will be $\pi_t$.
The upper index ' (``prime") will denote that the event changed any of the best prices, and the upper index $0$ that it did not. Abbreviations without the upper index ($\MO$, $\CA$, $\LO$) refer to both the price changing and the non-price changing event type. Every event is given a sign $\epsilon_t$ according to its expected long-term effect on the price -- the precise definitions are summarized in Table \ref{tab:eventtypes}. Note that the table also defines the gaps $\Delta_{\pi, \epsilon}$, which will be used later.
We will rely on indicator variables denoted as $I(\pi_{t}=\pi)$. This expression is equal to $1$ if the event at $t$ is of type $\pi$ and zero otherwise. We also use the notation $\ev{\cdot}$ to denote the time average of the quantity between the brackets. 

\begin{table*}[tbp]
    \begin{tabular}{|c|p{5cm}|p{3cm}|p{5.5cm}|}
	\hline
	$\pi$ & event definition & event sign definition & gap definition ($\Delta_{\pi, \epsilon}$) \\ \hline
	$\pi = \MOO$ & market order, volume $<$ outstanding volume at the best & $\epsilon = \pm 1$ for buy/sell market orders & $0$ \\ \hline
	$\pi = \MO'$ & market order, volume $\geq$ outstanding volume at the best & $\epsilon = \pm 1$ for buy/sell market orders & half of first gap behind the ask ($\epsilon=1$) or bid ($\epsilon=-1$) \\ \hline
	$\pi = \CAO$ & partial cancellation of the bid/ask queue & $\epsilon = \mp 1$ for buy/sell side cancellation & $0$ \\ \hline
	$\pi = \LOO$ & limit order at the current best bid/ask & $\epsilon = \pm 1$ for buy/sell limit orders & $0$ \\ \hline
	$\pi = \CA'$ & complete cancellation of the best bid/ask & $\epsilon = \mp 1$ for buy/sell side cancellation & half of first gap behind the ask ($\epsilon=1$) or bid ($\epsilon=-1$) \\ \hline
	$\pi = \LO'$ & limit order inside the spread & $\epsilon = \pm 1$ for buy/sell limit order & half distance of limit order from the earlier best quote on the same side \\ \hline
    \end{tabular}
    \caption{Summary of the $6$ possible event types, the corresponding definitions of the event signs and gaps.}
    \label{tab:eventtypes}
\end{table*}

Let us now define the response of the price to different types of orders. The average behavior of price after events of a particular type $\pi$ defines the corresponding \emph{response function} (or average impact function):
\begin{equation}
    {\cal R}_\pi(\ell) = \ev{(p_{t+\ell}-p_t) \cdot \epsilon_t |\pi_t = \pi}.
    \label{eq:Rpi}
\end{equation}
This is a correlation function between $\epsilon_t I(\pi_t = \pi)$ at time $t$ and the price change from $t$ to $t+\ell$, normalized by the stationary probability of the event $\pi$, denoted as $P(\pi) = \ev{I(\pi_t = \pi)}$. This normalized response function gives the expected directional price change after an event $\pi$. Its behavior for all $\pi$'s is shown in Fig. \ref{fig:RG}(left). Tautologically, ${\cal R}_\pi(\ell=1) >0$ for price changing events and ${\cal R}_\pi(\ell=1)=0$ for the others. Empirically, all types of events lead, on average, to a price change in the expected direction, i.e. ${\cal R}_\pi(\ell) > 0$.

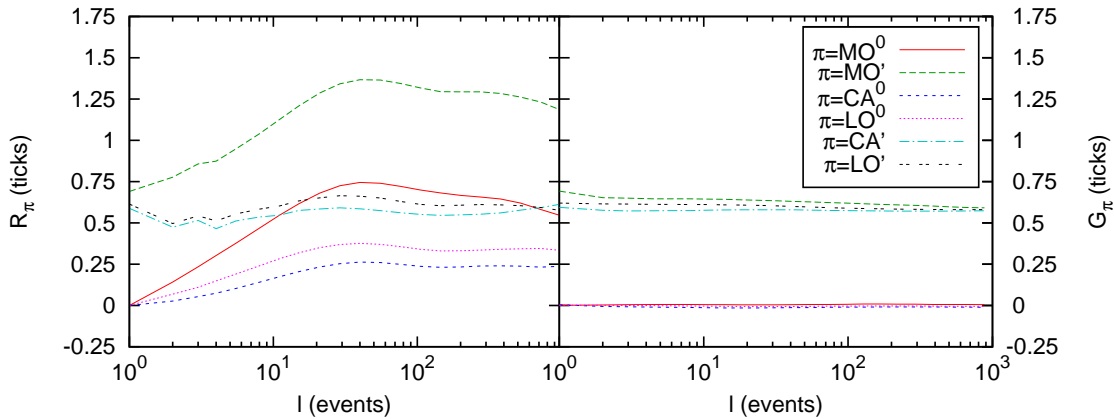
\addwidefig{RG}{./}{RG}{{\it (left)}: The response function $\R_\pi(\ell)$ and {\it (right)}: the bare impact function $G_\pi(\ell)$ for the TIM, for the data described in Sec. \ref{sec:data}. 
The curves are labeled according to $\pi$ in the legend. Note that $G_\pi(\ell)$ is to a first approximation independent of $\ell$ for all $\pi$'s. However, the small variations are real and 
important to reproduce the correct price diffusion curve $D(\ell)$.} 

We will also need the ``return'' response function, upgrading the quantity ${\cal S}(\ell)$ defined above to a matrix:
\begin{equation}
    {\cal S}_{\pi_1,\pi_2}(\ell) = \ev{I(\pi_{t+\ell}=\pi_2) \cdot r_{t+\ell} \cdot \epsilon_t |\pi_t = \pi_1}.
    \label{eq:Spi}
\end{equation}
Clearly, as an exact identity:
\begin{equation}
    {\cal R}_\pi(\ell) = \sum_{n=0}^{\ell-1}\sum_{\pi'} {\cal S}_{\pi,\pi'}(n).
\end{equation}
Similarly, the signed-event correlation function is defined as:
\begin{equation}
    C_{\pi_1, \pi_2}(\ell) = \frac{\ev{I(\pi_{t}=\pi_1)\epsilon_tI(\pi_{t+\ell}=\pi_2)\epsilon_{t+\ell}}}{P(\pi_1)P(\pi_2)}.
    \label{eq:Cpi1pi2}
\end{equation}
Our convention is that the first index corresponds to the first event in chronological order. Note that in general, there are no reasons to expect time reversal symmetry, which would impose $C_{\pi_1, \pi_2}(\ell)=C_{\pi_2, \pi_1}(\ell)$. If one has $N$ event types, altogether there are $N^2$ of these event-event correlation and return response functions.

\subsection{The transient impact model (TIM)}

Let us write down the natural generalization of the above transient impact model (TIM), embodied by Eq. \eqref{eq:ptMO}, to the many-event case. We now envisage that each event type $\pi$ has a ``bare'' 
time dependent impact given by $G_\pi(\ell)$, such that the price reads:
\begin{equation}
    p_t = \sum_{t'<t} \left[G_{\pi_t'}(t-t') \epsilon_{t'} + \eta_{t'}\right] + p_{-\infty},
    \label{eq:pt}
\end{equation}
where one selects for each $t'$, the propagator $G_{\pi_t'}$ corresponding to the particular event type at that time. After straightforward calculations, the response function \eqref{eq:Rpi} 
can be expressed as
\begin{eqnarray}
    \R_{\pi_1}(\ell+1)-\R_{\pi_1}(\ell) = \sum_{\pi_2} P(\pi_2) \left[ G_{\pi_2}(0) C_{\pi_1, \pi_2}(\ell) + \sum_{n>-\ell} \left[G_{\pi_2}(\ell+n+1)
    - G_{\pi_2}(\ell+n)\right] C_{\pi_1, \pi_2}(n) \right].
    \label{eq:undresspi}
\end{eqnarray}
This is a direct extension of Eq. \eqref{eq:RCG}. One can invert the system of equations \eqref{eq:undresspi}, to evaluate the unobservable $G_\pi$'s in terms of the observable $R_\pi$'s and $C_{\pi_1, \pi_2}$'s -- see Fig. \ref{fig:RG}(right).
Note that we formulate the problem here in terms of the ``derivatives'' of the $G_\pi$'s since, as mentioned above, it is numerically much more stable to introduce new variables for the increments of $\R$ and $G$ and solve \eqref{eq:undresspi} in terms of those.  

Once this is known, one can explicitly compute the lag dependent diffusion constant 
$D(\ell)=\ev{(p_{t+\ell}-p_t)^2}$ in terms of the $G$'s and the $C$'s, generalizing the corresponding result obtained in Ref. \cite{bouchaud.subtle}: see Appendix \ref{app:expr}.

\subsection{The history-dependent impact model (HDIM)}

Now, if one wants to generalize the history dependent impact model (HDIM), Eq. \eqref{eq:ptMO2}, to the many-event case, one immediately realizes that the impact at time $t$ depends on the type of
event one is considering. For one thing, the instantaneous impact of price non-changing events is trivially zero. For price changing events, $\pi'=\MO',\LO',\CA'$, the instantaneous impact is given by:
\begin{equation}\label{rt}
    r_t = \epsilon_{t} \Delta_{t}.
\end{equation}
Here $\Delta$ depends on the type of price changing event $\pi'$ that happens then, and possibly also on the sign of that event. For example, 
if $\pi_t=\MO'$ and $\epsilon_{t}=-1$ this means that at a sell market order executed the total volume at the bid. The midquote price change is $-\Delta_t$, which usually means that the
second best level was at $b_{t}-2\Delta_t$, where $b_{t}$ is the bid price before the event. The factor $2$ is necessary, because the ask
did not change, and the impact is defined by the change of the midquote. Hence $\Delta$'s for $\MO'$s (and similarly ${\CA'}$'s) correspond to half of the gap between the first 
and the second best quote \emph{just before} the level was removed (see also Ref. \cite{farmer.whatreally}). Another example when $\pi=\LO'$ and $\epsilon_{t}=-1$. This means that at $t$ 
a sell limit order was placed inside the spread. The midquote price change is $-\Delta$, which means that the limit order was placed at $a_{t}-2\Delta_t$, where $a_{t}$ 
is the ask price. Thus $\Delta$ for $\LO'$'s correspond to half of the gap between the first and the second best quote \emph{right after} the limit order was placed. In the following we will call the $\Delta$'s \emph{gaps}. For large tick stocks, these non-zero $\Delta$'s are most of the time equal to half a tick, and only very weakly fluctuating. For small tick stocks, substantial fluctuations of the spread, and of the gaps behind the best quotes, can take place. The generalization of Eq. \eqref{eq:ptMO2} precisely attempts to capture these gap fluctuations, which are affected by the flow of past events.  If we assume that the whole dynamical process is statistically invariant under exchanging buy orders with sell orders ($\epsilon \Rightarrow
- \epsilon$) and bids with asks, the dependence on the current non zero gaps on the past order flow can only include an even number of $\epsilon$'s. Therefore the lowest order model for 
non-zero gaps (including a constant term and a term quadratic in $\epsilon$'s) is:
\begin{equation}
\label{eq:delta}
\left. \Delta_{t} \right|_{\pi_t=\pi';\epsilon_t=\epsilon}  = \DeltaR_{\pi'} +  \sum_{t_1<t} \kappa_{\pi_{1},\pi'}(t-t_1)\left[ \epsilon_{t_1} \epsilon - C_{\pi_1, \pi'}(t-t_1) \right] + \eta'_t,
\end{equation}
where $\kappa_{\pi_1,\pi'}$ are kernels that model the dependence of the gaps on the past order flow (note that $\kappa_{\pi_1,\pi'}$ is a $6 \times 3$ matrix) 
and $\DeltaR_{\pi'}$ are the average realized gaps, defined as $\ev{\Delta_t|\pi_{t}=\pi'}$, since the average of the second term in the right hand side is identically zero. 
Note that the last term, equal to $\sum_{n>0}\sum_{\pi} P(\pi) \kappa_{\pi, \pi'}(n) C_{\pi, \pi'}(n)$, was not explicitly included in our previous analysis,
Ref. \cite{eisler.long}. However, typical values are less than $1\%$ of the average realized gap, and therefore negligible in practice. We set it to zero in the following.

Eq. \eqref{eq:delta}, combined with the definition of the return at time $t$, Eq. \eqref{rt}, leads to our generalization of the HDIM, Eq. \eqref{eq:ptMO2}:
\begin{equation}
\label{eq:finalmodel}
r_t = \DeltaR_{\pi_{t}} \epsilon_t +   \sum_{t_1<t} \kappa_{\pi_{t_1},\pi_t}(t-t_1) \epsilon_{t_1} + \eta_t.
\end{equation}
It is interesting to compare the above equation with its analogue for the transient impact model, which reads (after Eq. \eqref{eq:pt}):
\begin{equation}
r_t = G_{\pi_{t}}(1) \epsilon_t +   \sum_{t_1<t} \left[G_{\pi_{t_1}}(t-t_1+1)-G_{\pi_{t_1}}(t-t_1)\right] \epsilon_{t_1} + \eta_t.
\end{equation}
The two models can only be equivalent if:
\begin{equation}
\kappa_{\pi_{t_1},\pi}(\ell) \equiv G_{\pi_{t_1}}(\ell+1) - G_{\pi_{t_1}}(\ell), \qquad \forall \pi,
\label{eq:kappaequiv}
\end{equation}
which means that the ``influence matrix'' $\kappa_{\pi_1,\pi}$ has a much constrained structure, which has no reason to be optimal. It is also {\it a priori} inconsistent since the TIM 
leads to a non zero price move even if $\pi$ is a non price-changing event, since Eq. \eqref{eq:kappaequiv} is valid for all event types $\pi$. This is a major conceptual drawback of the 
TIM framework (although, as we will see below, the model fares quite well at reproducing the price diffusion curve).

The matrix $\kappa_{\pi_1,\pi_2}$ can in principle be determined from the empirical knowledge of the response matrices ${\cal S}_{\pi_1,\pi_2}$, since:  
\begin{eqnarray}
    \frac{1}{P(\pi_2)} {\cal S}_{\pi_1,\pi_2}(\ell) &=& \frac{1}{P(\pi_2)} \ev{I(\pi_{t+\ell}=\pi_2) \cdot r_{t+\ell} \cdot \epsilon_t |\pi_t = \pi_1}
    = \DeltaR_{\pi_2} C_{\pi_1,\pi_2}(\ell) 
   \\ \nonumber
   &+& \sum_{t' < t + \ell} \sum_\pi \kappa_{\pi,\pi_2}(t+\ell-t') \ev{I(\pi_{t'}=\pi)\epsilon_{t'} I(\pi_{t}=\pi_1)\epsilon_t | \pi_{t+\ell}=\pi_2}.
\end{eqnarray}
Note however that the last term includes a three-body correlation function which is not very convenient to estimate. At this stage and below, we need to make some approximation to estimate higher
order correlations. We assume that all three- and four-body correlation functions can be factorized in terms of two-body correlation functions, as if the variables were Gaussian. This allows to extract $\kappa_{\pi_1,\pi_2}$ from a numerically convenient expression, used in \cite{eisler.long}:
\begin{equation}
\label{eq:calib}
  \frac{1}{P(\pi_2)} {\cal S}_{\pi_1,\pi_2}(\ell) = \DeltaR_{\pi_2} C_{\pi_1,\pi_2}(\ell) + \sum_{n > - \ell} \sum_\pi \kappa_{\pi,\pi_2}(n+\ell) C_{\pi_1, \pi} (n). 
\end{equation}
Knowing the $\kappa_{\pi_1,\pi_2}$'s and using the same factorization approximation, one can finally estimate the price diffusion constant, given in Appendix \ref{app:expr}.
Although the factorization approximation used to obtain the diffusion constant looks somewhat arbitrary, we find that it is extremely precise when applied to the diffusion curve. 

For large tick stocks, the gaps hardly vary with time and are all equal to 1 tick. In other
words $\kappa_{\pi_1,\pi} \approx 0$ and the model simplifies enormously, since now $G_\pi(\ell) \equiv \Delta^\mathrm{R}_{\pi}$. In this limit, one therefore finds:
\begin{eqnarray}
R_{\pi}(\ell) = \ev{(p_{t+\ell}-p_t) \cdot \epsilon_t |\pi_t = \pi} \approx \Delta^\mathrm{R}_{\pi} + \sum_{0 < t'< \ell} \sum_{\pi_1} \Delta^\mathrm{R}_{\pi_1} P(\pi_1) C_{\pi, \pi_1}(t'),
    \label{eq:rmean}
\end{eqnarray}
which means that the total price response to some event can be understood as its own impact (lag zero), plus the sum of the biases in the course of future events, conditional to this initial event. These biases are multiplied by the average price change $\Delta^\mathrm{R}$ that these induced future events cause. Within the same model, the volatility reads:
\begin{eqnarray}
    D(\ell) = \ev{(p_{t+\ell}-p_t)^2} \approx \sum_{0\leq t', t''<\ell} \sum_{\pi_1} \sum_{\pi_2} P(\pi_1) P(\pi_2) 
    C_{\pi_1, \pi_2}(t'-t'') \Delta^\mathrm{R}_{\pi_1} \Delta^\mathrm{R}_{\pi_2}.
    \label{eq:vol}
\end{eqnarray}

For small ticks, on the other hand, gaps do fluctuate and react to the past order flow; the influence matrix $\kappa_{\pi_1,\pi'}$ describes how the past order flow affects the current gaps. If $\kappa_{\pi_1,\pi'}(\ell)$ is positive, it means that an event of type $\pi_1$ (price changing or not) tends to {\it increase} 
the gaps (i.e. reduce the liquidity) for a later price changing event $\pi'$ in the same direction, and decrease the gap if the sign of the event $\pi'$ is opposite to that of $\pi_1$.

\section{Model calibration and empirical tests}

\subsection{Data}

\label{sec:data}

We have tested the above ideas on a set of data made of $14$ randomly selected liquid stocks traded on the NASDAQ during the period 03/03/2008 -- 19/05/2008, a total of $53$ trading days (see Ref. \cite{eisler.long} for a detailed presentation of these stocks and summary statistics). In order to reduce the effects of the intraday spread and liquidity variations we exclude the first 30 and the last 40 minutes of the trading days. The particular choice of market is not very important, many of our results were also verified on other markets, such as CME Futures, US Treasury Bonds and stocks traded at the London Stock Exchange. 

Our sample of stocks can be divided into two groups: large tick and small tick stocks. Large tick stocks are such that the bid-ask spread is almost always equal to one tick, whereas small tick stocks have spreads that are typically a few ticks. 
The behavior of the two groups is quite different, for example, the events which change the best price have a relatively low probability for large tick stocks (about $3\%$ altogether), but not for small tick stocks (up to $40\%$).
Note that there is a number of stocks with intermediate tick sizes, which to some extent possess the characteristics of both groups. Technically, they can be treated in exactly the same way as small tick stocks, and all our results remain valid. However, for the clarity of presentation, we will not consider them explicitly.

As explained above, we restrict ourselves to events that modify the bid or ask price, or the volume quoted at these prices. Events deeper in the order book are unobserved and will not be described: 
although they do not have an immediate effect on the best quotes, our description is still incomplete. Furthermore, we note that the stocks we are dealing with are traded on multiple platforms.
This may account for some of the residual discrepancies reported below.

Since we consider 6 types of events, there are 6 response functions ${\cal R}_\pi$ and propagators $G_\pi$, 36 correlation functions $C_{\pi_1,\pi_2}$. However, since the 
return response functions ${\cal S}_{\pi_1,\pi_2}$ and the influence kernels $\kappa_{\pi_2, \pi_1}$ are non zero only when the second event $\pi_2$ is a price changing event, there are only $3 \times 6 =18$ of them. 

\subsection{The case of large ticks}

As explained in the previous section, the case of large ticks is quite simple since the gap fluctuation term of HDIM can be neglected altogether. As shown in Ref. \cite{eisler.long}, the predictions given
by Eqs. \eqref{eq:rmean} and \eqref{eq:vol} are in very good agreement with the empirical determination of the ${\cal R}_\pi(\ell)$ and the price diffusion $D(\ell)$. Small remaining discrepancies can indeed be accounted for by adding the gap fluctuation contribution, of the order of a few percents.  

The temporary impact model, on the other hand, is not well adapted to describe large tick stocks, for the following reason: when Eq. \eqref{eq:undresspi} is used to extract $G_\pi(\ell)$ from the data, small numerical errors may lead to some spurious time dependence. But as far as $D(\ell)$ is concerned, any small variation of $G_\pi$ is amplified through the second term of Eq. \eqref{eq:Dell} which is an infinite sum of positive terms. As noted in Ref. \cite{eisler.long}, this leads to large discrepancies between the predicted $D(\ell)$ and its empirical determination. At any rate, one should clearly 
favor the calibration of $G_\pi(\ell)$ using Eq. \eqref{eq:undresspi} rather than the analogue of Eq. \eqref{eq:RCG}.

\subsection{The case of small ticks}

The case of small ticks is much more interesting, since in this case the role of gap fluctuations is crucial, and is a priori a stringent test for the two models on stage. 

\subsubsection{TIM} 

Within the temporary impact model, the response functions ${\cal R}_\pi(\ell)$ are tautologically accounted for, since they are used to calibrate the propagators $G_\pi(\ell)$ using  Eq. \eqref{eq:undresspi}. Once the $G_\pi(\ell)$'s are
known (see Fig. \ref{fig:RG}, where ${\cal R}_\pi(\ell)$ and $G_\pi(\ell)$ are shown for small tick stocks), one can compute the time dependent diffusion coefficient $D(\ell)/\ell$, and compare with empirical data. This is shown in Fig. \ref{fig:undressY_voltest}(left). Note that we calibrate the $G_\pi(\ell)$ for each stock separately, compute $D(\ell)/\ell$ in each case, and then average the results over all stocks. The agreement is surprisingly good for long times, while for shorter times the model underestimates price fluctuations, which is expected since the model does not allow for 
high frequency fluctuations. 
We also show the prediction based on the constant gap approximation, $\kappa_{\pi_1,\pi_2} \equiv 0$. Although Fig. \ref{fig:RG} suggests that this is an acceptable assumption, we see that 
$D(\ell)$ is overestimated. As will be argued below, gaps do adapt to past order flow, and the net effect of the gap dynamics is to reduce the price volatility. 
We finally note that calibrating the $G_\pi(\ell)$ on the response functions directly (and not on their derivatives), as was done in \cite{eisler.long}, leads to much poorer results for the diffusion coefficient $D(\ell)/\ell$. 

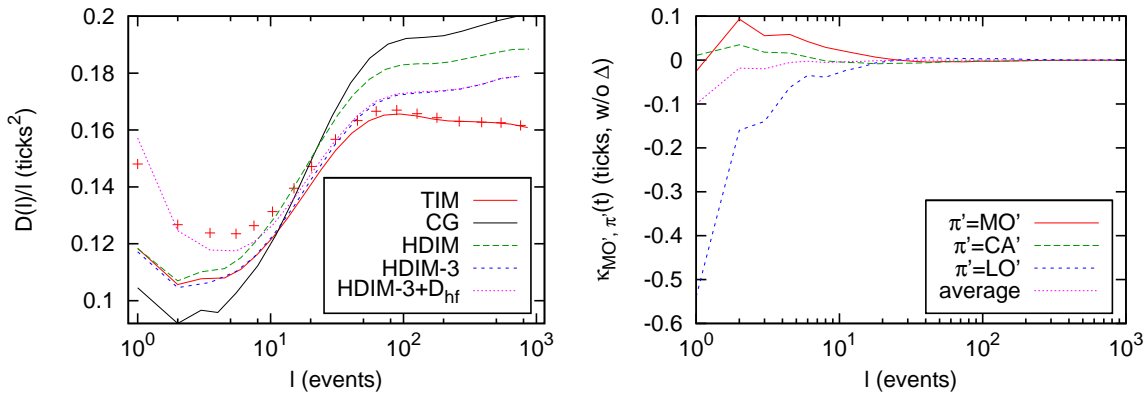
\addtwofigs{undressY_voltest}{./}{vol_test_single}{undressY_components_paper_smalltick}{{\it (left)} $D(\ell)/\ell$ and its approximations. Crosses correspond to the data. 
The constant gap (CG) model corresponds to $G_\pi(\ell) \equiv G_\pi(1)$. TIM corresponds to the temporary impact model calibrated on returns. The curve for HDIM uses the approximate calibration of $\kappa$'s, HDIM-3 is taking $3$ times the $\kappa$'s as from the calibration. We also indicate HDIM-3 with adding the constant $D_\mathrm{hf}=0.04$. Note that the vertical scale is different from Fig. 1, since in the time clock is different in the two cases (all events vs. trades in Fig. 1). {\it (right)} Comparison of the three non zero $\kappa_{\MO', \pi'}(\ell)$ with their average over $\pi'$. Note that 
$\kappa_{\MO', \MO'} < 0$: after an $\MO'$ event, gaps on the same side are on average smaller.}


\subsubsection{HDIM} 

We now turn to the history-dependent impact model. As explained above, we determine the influence kernels $\kappa_{\pi_1, \pi_2}(\ell)$ using Eq. \eqref{eq:calib}. We plot in Fig. \ref{fig:GY_single} the resulting 
``integrated impact'' on the future gaps of all 6 $\pi_1$ events, which we define as\footnote{Note that this definition is compatible with the one given in Ref. \cite{eisler.long}, because of a slight change in the
interpretation in the $\kappa$ kernels here.}:
\begin{equation}
\label{eq:gstarkappa}
\delta G^*_{\pi}(\ell) = \sum_{n=1}^{\ell - 1} \sum_{\pi'} P(\pi') \kappa_{\pi, \pi'}(n),
\end{equation}
As explained in Ref. \cite{eisler.long}, $\delta G^*_{\pi}(\ell)$ captures the contribution of the gap ``compressibility'' to the impact of an event of type $\pi$ up to a time lag $\ell$, leaving the 
sequence of events unchanged. If $\kappa_{\pi, \pi'}(n)$ were independent of $\pi'$, as postulated in the TIM, one would have $\delta G^*_{\pi}(\ell)=G_\pi(\ell)-G_\pi(1)$ as an identity. The agreement turns out to be excellent (see Fig. \ref{fig:GY_single}), which was not guaranteed a priori since the HDIM is calibrated on a much larger set of correlation functions.

However, this does not mean that  $\kappa_{\pi, \pi'}(n)$ are necessarily independent of $\pi'$. To illustrate the point that Eq. \eqref{eq:kappaequiv} is too restrictive, Fig. \ref{fig:undressY_voltest}(right) compares the three  $\kappa_{\MO',\pi'}$s, which are clearly different from one another. Note that the average over $\pi'$ is negative, meaning that $\MO'$ events tend to ``harden'' the book (i.e. after an $\MO'$ event, gaps on the same side are on average smaller). This is true for all price changing events, while (perhaps surprisingly) small market orders $\MOO$ ``soften" the book: $\delta G^*_{\MOO}$ is positive and gaps tend to grow. Queue fluctuations ($\CAO$ and $\LOO$) seem less important, but for small ticks these types of events also harden the book. Note finally that for large ticks $\delta G^*$'s are found to be about two orders of magnitude smaller, which confirms that gap fluctuations can be neglected in that case.


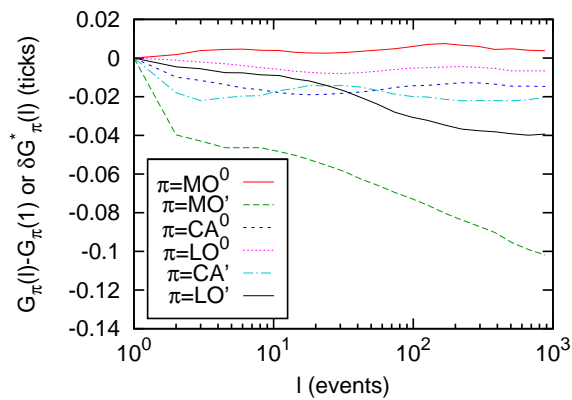
\addsmallfig{GY_single}{./}{GY_single}{The integrated impact on the future gaps $\delta G^*_\pi(\ell)$ in the HDIM estimated via Eq. \eqref{eq:kappaequiv}. The results are indistinguishable from $G_\pi(\ell)-G_\pi(1)$ calculated for the TIM. The curves are labeled according to $\pi$ in the legend.}

Now, Eq. \eqref{eq:calib} relies on the factorization of a three-point correlation function and is not exact, so there is no guarantee that the response functions ${\cal R}_\pi(\ell)$ are exactly reproduced using this calibration method. In order to check this approximation, we have simulated an artificial market dynamics where the price evolves according to Eq. \eqref{eq:finalmodel}, 
with the true (historical) sequence of signs and events and $\eta_t=0$. The kernels $\kappa_{\pi_1, \pi_2}$ are calibrated using Eq. \eqref{eq:calib}. This leads to the predictions shown as 
dashed lines in  Fig. \ref{fig:redressY_simu_smalltick_tweak}. The agreement can be much improved by simply multiplying all $\kappa$'s by a factor $3$, see  Fig. \ref{fig:redressY_simu_smalltick_tweak}. 
Of course, some discrepancies remain and one should use the historical simulation systematically to determine the optimal $\kappa$'s. This is, however, numerically much heavier and an improved analytical approximation of the three-point correlation function, that would allow a more accurate workable calibration, would be welcome.

Finally, we computed $D(\ell)$ for the HDIM using \eqref{eq:Dellhybrid2} in the Appendix. Here again, we have tested the quality of the factorization approximation using the same historical simulation. In this case, the $D(\ell)$ curve is indistinguishable from its approximation, so any discrepancy between the data and formula \eqref{eq:Dellhybrid2} cannot be blamed on its approximate nature, but rather on an inadequate calibration of the $\kappa$'s. 

The result is given in Fig. \ref{fig:undressY_voltest}(left) together with the previous theoretical predictions and the empirical data.\footnote{The $D(\ell)$-HDIM shown here is indistinguishable from the one appearing in Fig. 16 of Ref. \cite{eisler.long}.} With the naive calibration the HDIM turns out to be worse than the TIM for large lags: it overestimates $D(\ell)$ by $15 \%$ or so. Increasing the $\kappa$'s by a factor $3$ again greatly improves the fit but part of the discrepancy remains. For small lags, one needs to add a constant contribution $D_\mathrm{hf} \approx 0.04\mathrm{\ ticks\ squared}$ to match the data.\footnote{This contribution accounts for high frequency ``noise'' in the data that the model is not 
able to reproduce, as, for example, sequences of placement and cancellation of the same limit order inside the gap.} The HDIM produces a significant improvement over the constant gap model, because it explicitly includes the effect of gap fluctuations. However, since the calibration procedure relies on an approximation, we do not reproduce the response functions exactly. Hence the better founded model (HDIM) fares worse in practice than a model with theoretical inconsistencies (TIM). As noted above, a better calibration procedure for the $\kappa$'s could improve the situation. 

At any rate, numerical discrepancies should be expected regardless of the fitting procedure, since we have neglected several effects,
which must be present. These include (i) all volume dependence, (ii) unobserved events deeper in the book and on other platforms and (iii) higher order, non-linear contributions to model history dependence. On the last point, we note that based on symmetry arguments, the gap fluctuation term may include higher order terms of the form:
\begin{equation}
\sum_{t_1,t_2,t_3 <t'}\kappa_{\pi_{t_1},\pi_{t_2},\pi_{t_3};\pi_{t'}}(t'-t_1,t'-t_2,t'-t_3)\epsilon_{t_1} \epsilon_{t_2} \epsilon_{t_3}\epsilon_{t'},
\end{equation}
or with a larger (even) number of $\epsilon$'s. The presence of a four $\epsilon$ term is in fact suggested by the data shown in Fig. 13 of Ref. \cite{eisler.long}, and also by more recent analysis \cite{toth.unwritten}. 
It would be interesting to study these effects in detail, and understand their impact on price diffusion. 

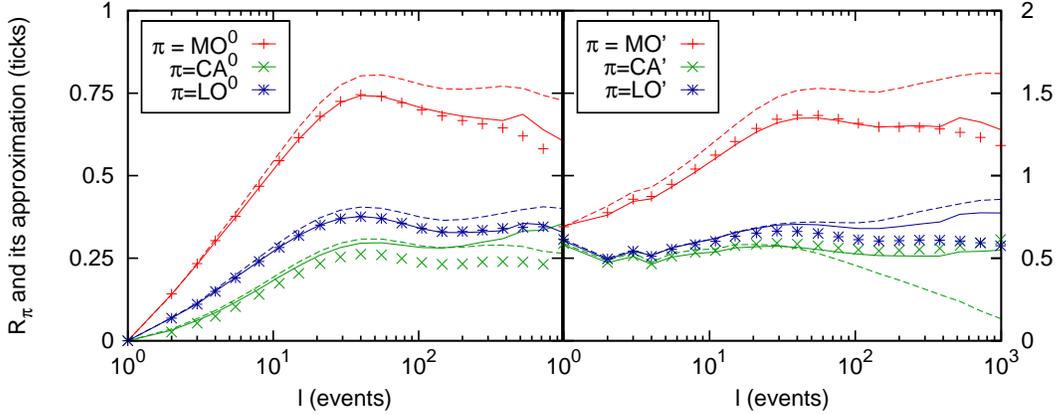
\addwidefig{redressY_simu_smalltick_tweak}{./}{redressY_simu_smalltick_tweak}{$\R_\pi(\ell)$ and their approximation with the HDIM. Symbols correspond to data, they are perfectly in line with the model prediction under the assumption that approximation \eqref{eq:calib} is correct. The dashed lines correspond to the response function of an actual simulation of the model with the $\kappa$'s calibrated via Eq. \eqref{eq:calib}. The solid lines correspond to the simulation if we increase all calibrated $\kappa$'s by a factor $3$ (HDIM-3). The colors vary according to $\pi$ as shown in the legend.}

\section{Conclusion}

Let us summarize what we have tried to accomplish in the present paper. Our aim was to provide a general framework to describe the impact of different events in the order book, in a way that is flexible enough to deal with any classification of these events (provided this classification makes sense).\footnote{See Ref. \cite{toth.brokerage} for an application of this method to orders 
with brokerage codes.} We have specifically considered market orders, limit orders and 
cancellation at the best quotes, further subdividing each category into price-changing and non price-changing events, giving a total of 6 types. In trying to generalize previous work, which focused 
on the impact of market orders only, we have discovered that two different models can be envisaged. These are equivalent when only a single event type, market orders regardless of their aggressivity, are taken into account. One model posits that each event type
has a temporary impact (TIM), whereas the other assumes that only price-changing events have a direct impact, which is itself modulated by the past history of all events, a model we called ``history
dependent impact'' (HDIM). 

The TIM is a natural extension of Hasbrouck's VAR model to a multi-event setting: one writes a Vector Autoregression model for the return at time $t$ in terms of {\it all} 
signed past events, but neglects the direct influence of past returns themselves (although these would be easy to include if needed). We have discussed the fact that TIMs are, strictly speaking, inconsistent since they assign a non-zero immediate impact to non price-changing events. Still, provided the model is correctly calibrated using returns (see Eq. \eqref{eq:undresspi}), we find that the TIM 
framework allows one to reproduce the price diffusion pattern surprisingly accurately. 

The HDIM family can also be thought of as a VAR model, although  one now distinguishes between different types of event-induced returns before regressing them on past events. The HDIM is interesting because it gives a very appealing interpretation of the price changing process
in terms of history dependent ``gaps'' that determine the amplitude of the price jump if a certain type of price-changing event takes place. We have in particular defined a lag-dependent, $6 \times 3$ ``influence matrix'' (called $\kappa_{\pi,\pi'}$ in the text), which tells us how much, on average, an event of type $\pi$ affects the immediate impact of a $\pi'$ price-changing event of the same sign 
in the future. 

The HDIM therefore envisages the dynamics of prices as consisting of three processes: instantaneous jumps due to events, events inducing further events and thereby affecting the future jump {\it probabilities} (described by the correlation between events), and events exerting pressure on the gaps behind the best price and thereby affecting the future jump {\it sizes} (described by the 
$\kappa$'s). By describing this third effect with a linear regression process, we came up with the explicit model \eqref{eq:finalmodel}, that can be calibrated on empirical data provided some factorization approximation is made (which unfortunately turns out not to be very accurate, calling for further work on this matter). This allows one to measure the influence matrix $\kappa$ and its lag dependence. We find in particular that price-changing events, such as 
aggressive market orders $\MO'$, tend to reduce the impact of later events of the same sign (i.e. a buy $\MO'$ following a buy $\MO'$) but increase the impact of later events of the opposite sign. 
As stressed in Refs. \cite{gerig.phd,bouchaud.review,lillo-farmer.qf}, this history dependent asymmetric liquidity is the dominant effect that mitigates persistent trends in prices which would otherwise be induced by the long-ranged correlation in the sign of market orders. 

In spite of these enticing features, we have found that the HDIM leads to a worse determination of the price diffusion properties than the TIM. The almost perfect agreement 
between the TIM prediction and empirical data is perhaps accidental, but it may also be that TIMs (that have less parameters) are numerically more robust than HDIMs. For HDIMs, a more accurate 
calibration procedure is needed. This could be achieved either by finding a better, workable approximation for the three-point correlation function, or by using a purely numerical approach based on a 
historical simulation of the HDIM. On the other hand, some effects have been explicitly neglected, such as the role of unobserved events deeper in the book and on other platforms, or possible non-linearities in the history dependence of gaps. It would be very interesting to investigate the relevance of these effects, and to come up either with a fully consistent version of HDIM, or with a convincing argument for why the TIM appears to be particularly successful. 

In any case, we hope that the intuitive and versatile framework that we proposed above, together with operational calibration procedures, will help making sense of the highly complex and intertwined sequences of events that take place in the order books, and allows one to build a comprehensive theory of price formation in electronic markets.       

\section*{Acknowledgments}

The authors are grateful to Emmanuel S\'eri\'e for his ideas on fitting impact kernels. They also thank Bence T\'oth for his critical reading and comments.

\appendix

\section{Expression of the price diffusion for the TIM and HDIM}
\label{app:expr}

We give here the rather ugly looking explicit expressions for the diffusion curve $D(\ell)$ in both models. For the TIM, one gets as an exact expression:
\begin{eqnarray}
  \label{eq:Dell}
  D(\ell) &=& D_0 \ell + \sum_{0\leq n < \ell}\sum_{\pi_1} G_{\pi_1}(\ell-n)^2 P(\pi_1) + 
  \sum_{n>0} \sum_{\pi_1} \left [G_{\pi_1}(\ell+n)-G_{\pi_1}(n) \right ]^2 P(\pi_1)  \nonumber \\ 
 &+& 2 \sum_{0\leq n < n' < \ell}\sum_{\pi_1, \pi_2} G_{\pi_1}(\ell-n)G_{\pi_2}(\ell-n')C_{\pi_1,\pi_2}(n'-n)  \nonumber \\
 &+& 2 \sum_{0 < n < n' < \ell}\sum_{\pi_1, \pi_2}\left [G_{\pi_1}(\ell+n)-G_{\pi_1}(n)\right ]\left [G_{\pi_2}(\ell+n')-G_{\pi_2}(n')\right ]C_{\pi_1,\pi_2}(n-n') \nonumber \\
 &+& 2 \sum_{0 \leq n < \ell} \sum_{n' > 0} \sum_{\pi_1, \pi_2} G_{\pi_1}(\ell-n)\left [G_{\pi_2}(\ell+n')-G_{\pi_2}(n')\right ]C_{\pi_2,\pi_1}(n'+n).
\end{eqnarray}
where $D_0$ is the variance of the noise term $\eta_t$. 

For the HDIM, on the other hand, one has to use a factorization approximation to compute 3- and 4-point correlation functions in terms of 2-point correlations. 
One can finally estimate the price diffusion constant, which is given by the following approximate equation \cite{eisler.long}:
\begin{eqnarray}
  D(\ell) = \ev{(p_{t+\ell}-p_t)^2} \approx D_0 \ell + \sum_{0\leq t', t''<\ell} \sum_{\pi_1} \sum_{\pi_2} P(\pi_1) P(\pi_2) 
  C_{\pi_1, \pi_2}(t'-t'') \Delta^\mathrm{R}_{\pi_1} \Delta^\mathrm{R}_{\pi_2} + \nonumber \\
  2 \sum_{-\ell < t < \ell} \sum_{\pi_2, \pi_3} \sum_{\tau > 0} (\ell - |t|) \DeltaR_{\pi_3} \kappa^+_{\pi_2, \pi_3} (\tau, t) C_{\pi_2, \pi_3}(t+\tau) P(\pi_2)P(\pi_3) + \nonumber \\
  \sum_{-\ell < t < \ell} \sum_{\pi_2, \pi_4}\sum_{\tau, \tau'>0} (\ell - |t|) \kappa^{++}_{\pi_2,\pi_4}(\tau, \tau',t)C_{\pi_2, \pi_4}(\tau-\tau'+t)P(\pi_2)P(\pi_4),
  \label{eq:Dellhybrid2}
\end{eqnarray}
where $D_0$ is again the variance of the noise $\eta_t$, 
\begin{eqnarray}
  \kappa^+_{\pi_2, \pi_3}(\tau, t) = \sum_{\pi_1} \kappa_{\pi_2, \pi_1}(\tau)[I(t=0)I(\pi_1=\pi_3)+I(t\not = 0)P(\pi_1)+
  I(t = -\tau) P(\pi_1) \Pi_{\pi_2 \pi_1} (\tau)],
\end{eqnarray}
and, for $t \geq 0$, 
\begin{eqnarray}
  \kappa^{++}_{\pi_2,\pi_4}(\tau, \tau', t) = \sum_{\pi_1, \pi_3} \kappa_{\pi_2, \pi_1}(\tau)\kappa_{\pi_4, \pi_3}(\tau')\{I(t=\tau')I(\pi_1=\pi_4)P(\pi_3) + \nonumber \\ I(t\not=\tau')P(\pi_1)P(\pi_3)[\Pi_{\pi_1, \pi_3}(t)+1]\},
\end{eqnarray}
whereas for $t<0$, we use $\kappa^{++}_{\pi_2,\pi_4}(\tau, \tau', -t) = \kappa^{++}_{\pi_4,\pi_2}(\tau', \tau, t)$. We also introduced a correlation function between event types as \cite{eisler.long}:
\begin{equation}
\Pi_{\pi_1,\pi_2}(\ell) = \frac{P(\pi_{t+\ell}=\pi_2|\pi_t=\pi_1)}{P(\pi_2)}-1 \equiv \frac{\ev{I(\pi_t=\pi_1)I(\pi_{t+\ell}=\pi_2)}}{P(\pi_1)P(\pi_2)}-1.
\end{equation}

\bibliography{limit}

\begin{thebibliography}{15}
\expandafter\ifx\csname natexlab\endcsname\relax\def\natexlab#1{#1}\fi
\expandafter\ifx\csname bibnamefont\endcsname\relax
  \def\bibnamefont#1{#1}\fi
\expandafter\ifx\csname bibfnamefont\endcsname\relax
  \def\bibfnamefont#1{#1}\fi
\expandafter\ifx\csname citenamefont\endcsname\relax
  \def\citenamefont#1{#1}\fi
\expandafter\ifx\csname url\endcsname\relax
  \def\url#1{\texttt{#1}}\fi
\expandafter\ifx\csname urlprefix\endcsname\relax\def\urlprefix{URL }\fi
\providecommand{\bibinfo}[2]{#2}
\providecommand{\eprint}[2][]{\url{#2}}

\bibitem[{\citenamefont{Hasbrouck}(2007)}]{hasbrouck.book}
\bibinfo{author}{\bibfnamefont{J.}~\bibnamefont{Hasbrouck}},
  \emph{\bibinfo{title}{Empirical Market Microstructure: The Institutions,
  Economics, and Econometrics of Securities Trading}}
  (\bibinfo{publisher}{Oxford University Press}, \bibinfo{year}{2007}).

\bibitem[{\citenamefont{Mike and Farmer}(2008)}]{mike.empirical}
\bibinfo{author}{\bibfnamefont{S.}~\bibnamefont{Mike}} \bibnamefont{and}
  \bibinfo{author}{\bibfnamefont{J.~D.} \bibnamefont{Farmer}},
  \bibinfo{journal}{Journal of Economic Dynamics and Control}
  \textbf{\bibinfo{volume}{32}}, \bibinfo{pages}{200} (\bibinfo{year}{2008}).

\bibitem[{\citenamefont{Bouchaud et~al.}(2004)\citenamefont{Bouchaud, Gefen,
  Potters, and Wyart}}]{bouchaud.subtle}
\bibinfo{author}{\bibfnamefont{J.-P.} \bibnamefont{Bouchaud}},
  \bibinfo{author}{\bibfnamefont{Y.}~\bibnamefont{Gefen}},
  \bibinfo{author}{\bibfnamefont{M.}~\bibnamefont{Potters}}, \bibnamefont{and}
  \bibinfo{author}{\bibfnamefont{M.}~\bibnamefont{Wyart}},
  \bibinfo{journal}{Quantitative Finance} \textbf{\bibinfo{volume}{4}},
  \bibinfo{pages}{176} (\bibinfo{year}{2004}).

\bibitem[{\citenamefont{Bouchaud et~al.}(2006)\citenamefont{Bouchaud,
  Kockelkoren, and Potters}}]{bouchaud.molasses}
\bibinfo{author}{\bibfnamefont{J.-P.} \bibnamefont{Bouchaud}},
  \bibinfo{author}{\bibfnamefont{J.}~\bibnamefont{Kockelkoren}},
  \bibnamefont{and} \bibinfo{author}{\bibfnamefont{M.}~\bibnamefont{Potters}},
  \bibinfo{journal}{Quantitative Finance} \textbf{\bibinfo{volume}{6}},
  \bibinfo{pages}{115} (\bibinfo{year}{2006}).

\bibitem[{\citenamefont{Lyons}(2006)}]{lyons.book}
\bibinfo{author}{\bibfnamefont{R.~K.} \bibnamefont{Lyons}},
  \emph{\bibinfo{title}{The Microstructure Approach to Exchange Rates}}
  (\bibinfo{publisher}{MIT Press}, \bibinfo{year}{2006}).

\bibitem[{\citenamefont{Bouchaud et~al.}(2009)\citenamefont{Bouchaud, Farmer,
  and Lillo}}]{bouchaud.review}
\bibinfo{author}{\bibfnamefont{J.-P.} \bibnamefont{Bouchaud}},
  \bibinfo{author}{\bibfnamefont{J.~D.} \bibnamefont{Farmer}},
  \bibnamefont{and} \bibinfo{author}{\bibfnamefont{F.}~\bibnamefont{Lillo}},
  \emph{\bibinfo{title}{How markets slowly digest changes in supply and
  demand}}, \bibinfo{howpublished}{in Handbook of Financial Markets: Dynamics
  and Evolution, North-Holland, Elsevier} (\bibinfo{year}{2009}).

\bibitem[{\citenamefont{Weber and Rosenow}(2005)}]{weber.rosenow}
\bibinfo{author}{\bibfnamefont{P.}~\bibnamefont{Weber}} \bibnamefont{and}
  \bibinfo{author}{\bibfnamefont{B.}~\bibnamefont{Rosenow}},
  \bibinfo{journal}{Quantitative Finance} \textbf{\bibinfo{volume}{5}},
  \bibinfo{pages}{357} (\bibinfo{year}{2005}).

\bibitem[{\citenamefont{Gerig}(2008)}]{gerig.phd}
\bibinfo{author}{\bibfnamefont{A.}~\bibnamefont{Gerig}},
  \emph{\bibinfo{title}{A theory for market impact: How order flow affects
  stock price, \protect{PhD} thesis}} (\bibinfo{year}{2008}),
  \bibinfo{note}{arXiv:0804.3818}.

\bibitem[{\citenamefont{Farmer et~al.}(2006)\citenamefont{Farmer, Gerig, Lillo,
  and Mike}}]{lillo-farmer.qf}
\bibinfo{author}{\bibfnamefont{J.~D.} \bibnamefont{Farmer}},
  \bibinfo{author}{\bibfnamefont{A.}~\bibnamefont{Gerig}},
  \bibinfo{author}{\bibfnamefont{F.}~\bibnamefont{Lillo}}, \bibnamefont{and}
  \bibinfo{author}{\bibfnamefont{S.}~\bibnamefont{Mike}},
  \bibinfo{journal}{Quantitative Finance} \textbf{\bibinfo{volume}{6}},
  \bibinfo{pages}{107} (\bibinfo{year}{2006}).

\bibitem[{\citenamefont{Hautsch and Huang}(2009)}]{hautsch.limit}
\bibinfo{author}{\bibfnamefont{N.}~\bibnamefont{Hautsch}} \bibnamefont{and}
  \bibinfo{author}{\bibfnamefont{R.}~\bibnamefont{Huang}},
  \emph{\bibinfo{title}{The market impact of a limit order}}
  (\bibinfo{year}{2009}), \bibinfo{note}{working Paper 2009/23, Center for
  Financial Studies, Frankfurt am Main}.

\bibitem[{\citenamefont{Eisler et~al.}(2011)\citenamefont{Eisler, Bouchaud, and
  Kockelkoren}}]{eisler.long}
\bibinfo{author}{\bibfnamefont{Z.}~\bibnamefont{Eisler}},
  \bibinfo{author}{\bibfnamefont{J.-P.} \bibnamefont{Bouchaud}},
  \bibnamefont{and}
  \bibinfo{author}{\bibfnamefont{J.}~\bibnamefont{Kockelkoren}}
  (\bibinfo{year}{2011}), \bibinfo{note}{arXiv:0904.0900, to appear in
  Quantitative Finance}.

\bibitem[{\citenamefont{S\'eri\'e}(2010)}]{e.serie}
\bibinfo{author}{\bibfnamefont{E.}~\bibnamefont{S\'eri\'e}}
  (\bibinfo{year}{2010}), \bibinfo{note}{unpublished report, Capital Fund
  Management, Paris, France}.

\bibitem[{\citenamefont{T\'oth et~al.}(2011)\citenamefont{T\'oth, Eisler,
  Lillo, Bouchaud, Kockelkoren, and Farmer}}]{toth.brokerage}
\bibinfo{author}{\bibfnamefont{B.}~\bibnamefont{T\'oth}},
  \bibinfo{author}{\bibfnamefont{Z.}~\bibnamefont{Eisler}},
  \bibinfo{author}{\bibfnamefont{F.}~\bibnamefont{Lillo}},
  \bibinfo{author}{\bibfnamefont{J.-P.} \bibnamefont{Bouchaud}},
  \bibinfo{author}{\bibfnamefont{J.}~\bibnamefont{Kockelkoren}},
  \bibnamefont{and} \bibinfo{author}{\bibfnamefont{J.}~\bibnamefont{Farmer}}
  (\bibinfo{year}{2011}), \bibinfo{note}{arXiv:1104.0587}.

\bibitem[{\citenamefont{Farmer et~al.}(2004)\citenamefont{Farmer, Gillemot,
  Lillo, Mike, and Sen}}]{farmer.whatreally}
\bibinfo{author}{\bibfnamefont{J.~D.} \bibnamefont{Farmer}},
  \bibinfo{author}{\bibfnamefont{L.}~\bibnamefont{Gillemot}},
  \bibinfo{author}{\bibfnamefont{F.}~\bibnamefont{Lillo}},
  \bibinfo{author}{\bibfnamefont{S.}~\bibnamefont{Mike}}, \bibnamefont{and}
  \bibinfo{author}{\bibfnamefont{A.}~\bibnamefont{Sen}},
  \bibinfo{journal}{Quantitative Finance} \textbf{\bibinfo{volume}{4}},
  \bibinfo{pages}{383} (\bibinfo{year}{2004}).

\bibitem[{\citenamefont{T\'oth}(2011)}]{toth.unwritten}
\bibinfo{author}{\bibfnamefont{B.}~\bibnamefont{T\'oth}}
  (\bibinfo{year}{2011}), \bibinfo{note}{in preparation}.

\end{thebibliography}

\end{document}